%% file: Fast_3d_RT_Tomography.tex
\documentclass{article}
\usepackage{times}
\usepackage[intlimits]{amsmath}
\usepackage{amssymb}
\usepackage{url}
\usepackage{bm}
\usepackage{enumerate}
\usepackage{algorithmicx}
\usepackage[ruled]{algorithm}
\usepackage[]{algpseudocode}
\usepackage{paralist}
\usepackage{cite}
\usepackage{soul}

\usepackage{layouts}
\usepackage{booktabs}
\usepackage{colortbl}
\usepackage{endnotes}
\usepackage{xcolor}
\usepackage[normalem]{ulem}
\usepackage{import}
\usepackage{graphicx}
\usepackage{mathrsfs}
\usepackage{lipsum}
\usepackage{subcaption}
\DeclareCaptionSubType*{algorithm}

\DeclareCaptionLabelFormat{alglabel}{Alg.~#2}

\DeclareMathAlphabet\mathbfcal{OMS}{cmsy}{b}{n}
\newcommand{\Grad}[1]{\triangledown_{#1}}

\newcommand{\argmin}{\mathrm{arg}\min}

\newcommand{\curly}[1]{\left\{#1\right\}}

\newcommand{\yoavcomment}[1]{}
\renewcommand{\yoavcomment}[1]{#1} 

\newcommand{\diag}{\mathop{\mathrm{diag}}}

\newcommand{\xw}{\left({\bm x},{\bm \omega} \right)}
\newcommand{\operator}[1]{\mathbfcal{#1}}
\newcommand\authnote[1]{\textsuperscript{\normalfont#1}}

\providecommand\textsuperscript[1]{$^{#1}$}
\graphicspath{{./Figures/}}

\begin{document}


\title{An Efficient Approach for Optical Radiative Transfer Tomography 
using the Spherical Harmonics Discrete Ordinates Method}
\date{}
\author{Aviad Levis\authnote{1}, Yoav Y.~Schechner\authnote{1}, Amit Aides\authnote{1}, Anthony B.~Davis\authnote{2}}

\newcommand{\Addresses}{{
  \medskip
  \footnotesize

\textsc{1 Electrical Engineering Department, Technion - Israel Institute of Technology, Haifa 32000, Israel}\par\nopagebreak

  \medskip

   \textsc{2 Jet Propulsion Laboratory, California Institute of
  Technology, Pasadena, CA 91109, USA}\par\nopagebreak
}}
\maketitle

\Addresses

\begin{abstract}
This paper introduces a method to preform optical tomography, using 3D radiative transfer as the forward model. We use an iterative approach predicated on the Spherical Harmonics Discrete Ordinates Method (SHDOM) to solve the optimization problem in a scalable manner. We illustrate with an application in remote sensing of a cloudy atmosphere.
\end{abstract}


\section{Introduction}
\label{sec:intro}

Optical tomography is a 3D imaging technique that uses optical measurements on the boundary of a domain to find the spatial distribution of parameters within. It has numerous applications in bio-medical imaging. For a list of applications we refer the reader to \cite{bal2009inverse,arridge2009optical} and references therein.
In sharp contrast, operational remote sensing of clouds in the Earth's atmosphere using space-borne optical sensors is predicated on the classic plane-parallel medium geometry with horizontally uniform properties, hence the 1D radiative transfer equation (RTE); this reduces the problem to estimating a single cloud optical thickness, and  shifts the focus toward microphysical properties \cite{nakajima1990determination1,nakajima1991determination2,platnick2003modis}. Some of the biases in retrieved cloud properties caused by this sometimes very crude approximation of real cloud structure are documented in \cite{davis2010solar}.

Solving the inverse problem using the full 3D RTE as a forward model can be difficult and computationally demanding. For some applications, it is possible to use an approximate model. In dense media (mean free path small compared the overall size), with scattering dominating absorption, it is possible to use a diffusion approximation of the 3D RTE. This results in the inverse problem of {\em Diffuse Optical Tomography} (DOT) \cite{gibson2005recent,boas2001imaging,ntziachristos2000concurrent}. When the mean free path is large compared to the outer size of the medium, the measured radiant energy is dominated by direct and single scattered intensities. The resulting inverse problem is single scattering tomography \cite{arridge2009optical,florescu2009single,aides2013}. Other approximations and their derivations can be found in \cite{hancourant2008}. We propose to solve the inverse problem using the RTE, with neither single scattering, nor diffusion approximations. However, to make the inverse problem tractable, we derive an iterative optimization framework that uses different parts of the Spherical Harmonics Discrete Ordinates Method (SHDOM) 3D RTE solver \cite{evans1998}.


\section{Theoretical background}
\label{sec:theor-backgr}

\subsection{Radiative transfer}
\label{subsec:RTE}

\begin{figure}[t]
   \centering
    \def\svgwidth{\columnwidth}
    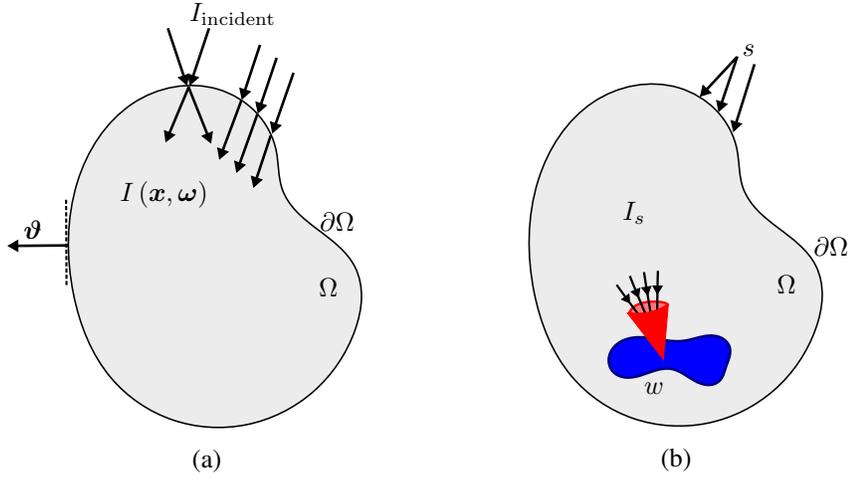
    \caption{(a) Domain and boundary conditions; (b) Aperture function. Blue marks the spatial support and red marks the angular support}
    \label{fig:domain}
\end{figure}

Our forward model is steady-state 3D radiative transfer. This model is used in passive imaging when source gating is sufficiently slow. The 3D RTE governs the transport of light through a medium with distribution of scattering and/or absorption. Consider a domain $\Omega$ having boundary $\partial \Omega$ whose outward facing normal is ${\bm \vartheta}$ (Fig.~\ref{fig:domain}a). The transport domain is indicated by position ${\bm x} \in \mathbb{R}^3$ and direction of propagation ${\bm \omega} \in \mathbb{S}^2$ (unit sphere). The radiation (light) field is $I_{\lambda}\xw$. The subscript $\lambda$ indicates wavelength dependency and will be omitted from here on, for simplicity. The boundary condition (Fig.~\ref{fig:domain}a) is
\begin{align}
I\xw = I_{\rm incident}\xw \qquad {\rm if} \quad {\bm \omega} \cdot {\bm \vartheta} < 0, \quad {\bm x} \in \partial \Omega,
\label{eq:BC}
\end{align}
where ${\bm \omega} \cdot {\bm \vartheta} < 0$ defines incoming radiation, ${\bm \vartheta}$ being the outgoing normal to $\partial\Omega$. The transport of light is formulated in terms of the following conservation law \cite{chandrasekhar1960,varadan1983}:
\begin{align}
   {\bm \omega} \cdot \Grad{} I \xw = \beta\left({\bm x}\right) \left[J\left({\bm x},{\bm \omega}\right) - I\xw\right] \quad {\bm x} \in \Omega,
   \label{eq:RTE}
\end{align}
 where ${\bm \beta}$ is the extinction coefficient. Here $J \xw$ is the {\em source function}. In the absence of emission within the medium, it is also known as the {\em in-scattering} term, since it accounts of an increase of radiation due to in-scattering. In this case, it is expressed as
\begin{align}
 J\xw = \frac{\varpi}{4\pi}\int_{\mathbb{S}^2} p\left({\bm \omega}\cdot{\bm \omega'}\right)I\left({\bm x},{\bm \omega'}\right) d {\bm \omega}',
 \label{eq:srcfun}
\end{align}
where $\varpi$ is the single scattering albedo and $p\left({\bm \omega}\cdot{\bm \omega'}\right)$\footnote{In principle, the phase function depends on both incoming and outgoing directions $\left({\bm \omega},{\bm \omega'}\right)$. However, dependency is often assumed to be solely on the scattering angle, equivalent to ${\bm \omega}\cdot{\bm \omega'}$.}  is the phase function. The phase function describes the probability of a photon traveling in direction ${\bm \omega}'$ into scatter to direction ${\bm \omega}$. The phase function satisfies a normalization condition
\begin{align}
 \frac{1}{4\pi}\int_{\mathbb{S}^2} p\left({\bm \omega}\cdot{\bm \omega'}\right)d{\bm \omega}' = 1.
 \label{eq:phasefunction}
\end{align}
Eqs.~(\ref{eq:RTE})--(\ref{eq:srcfun}), and the boundary condition (\ref{eq:BC}), define a complete radiative transfer forward model for an externally illuminated medium.

Integrating Eq.~(\ref{eq:RTE}) along a specific direction ${\bm \omega}$ results in the integral form of the 3D RTE \cite{chandrasekhar1960,davis20053d}
\begin{align}
I({\bm x},{\bm \omega}) &=  I_0 \exp{\left[-\int_{\bm x}^{{\bm x}_0}\beta\left({\bm r}\right)d{\bm r}\right]} \nonumber \\ & + \int_{\bm x}^{{\bm x}_0} J({\bm x}',{\bm \omega}) \beta\left({\bm x}'\right)\exp{\left[-\int_{{\bm x}}^{{\bm x}'}\beta\left({\bm r}\right)d{\bm r}\right]}d{\bm x}'  ,
\label{def:Tdef}
\end{align}
bearing in mind the expression of $J\xw$ in terms of $I\xw$ in (\ref{eq:srcfun}). Here ${\bm x}_0$ is a point on the boundary and $I_0$ holds the boundary condition (\ref{eq:BC}).
The operation we denote as
\begin{align}
\int_{{\bm x}}^{{\bm x}'} f({\bm r}) d{\bm r}
\label{def:intergration_operator}
\end{align}
is simply a 1D (line) integral over a field $f({\bm x})$ along the segment extending from ${\bm x}$ to ${\bm x}'$. Numerically, this is preformed by {\em back-projecting} (BP) a ray through the medium.

The expression for $I\xw$ in (\ref{def:Tdef}), for a given $J\xw$, is often referred to as the {up-wind sweep} of radiant energy sources. We adopted the {\em Spherical Harmonics Discrete Ordinates Method} (SHDOM)  \cite{evans1998,evans2005} for the forward 3D RT, which is available as open-source code. In SHDOM, (\ref{def:Tdef}) is indeed used in a straightforward post-processing stage for estimating the observed quantity $I\xw$ after a nontrivial numerical procedure is performed to obtain $J\xw$ throughout the grid. This staged approach is exploited further on. We now formalize our forward model in operator language.

\subsection{Operator Notation}
\label{subsec:opnotation}

\begin{table}
\begin{tabular}{l l l l}
\toprule[1.1pt]
	Operator & Symbol & Domain & Range \\ 	\hline
	Radiance forward mapping & $\operator{I} $ &  $\Theta$ - Extinction fields & $\mathcal{Z}$ - Radiance fields \\
	Measurement & $\operator{M} $ &  $\mathcal{Z}$ - Radiance fields & $\mathcal{Y}$ - Measurements\\
	Forward operator & $\operator{F} $ &  $\Theta$ - Extinction fields & $\mathcal{Y}$ - Measurements\\
	Scatter forward mapping & $\operator{J} $ &  $\Theta$ - Extinction fields & $\mathcal{V}$ - Scatter fields \\
	Transformation operator & $\operator{T} $ &  $\mathcal{V}$ - Scatter fields & $\mathcal{Z}$ - Radiance fields \\
\bottomrule[1.1pt]
\end {tabular}
\caption {Operator notation summary}
\label{tab:operators}
\end{table}

We follow the definitions of \cite{arridge2009optical} to formulate the forward model using operator notations (Table \ref{tab:operators} summarizes the notations). Denote by $\Theta$ the space of extinction fields over the domain $\Omega$. Let $\beta\left({\bm x}\right)\in \Theta$ be an extinction field over the domain $\Omega$.
Let $\mathcal{S}$ represent a set of radiation sources over $\partial \Omega$. For a given source $s\in\mathcal{S}$, denote $\mathcal{Z}_s$ as the set of all possible radiation fields that satisfy Eqs. (\ref{eq:BC},\ref{eq:RTE}) across all possible $\beta \in \Theta$. The set $\mathcal{Z}_s$ is infinite, since for each extinction field $\beta({\bm x})$ there is a corresponding radiation field $I_s = \operator{I}_s \left(  \beta \right)$. More generally the {\em  radiance forward mapping}, $ \operator{I}: \Theta \rightarrow \mathcal{Z}$, is a mapping from the optical parameter domain, to the radiance field range $\mathcal{Z} = \bigcup_s \mathcal{Z}_s$. Measurements are an operator $ \operator{M}: \mathcal{Z} \rightarrow \mathcal{Y}$, mapping the continuous function space to a vector space $\mathcal{Y}$. Consequently the forward operator  $\operator{F}$ is defined as
\begin{align}
 \operator{F}=\operator{M} \operator{I}:\Theta \rightarrow \mathcal{Y}.
 \label{def:forward}
\end{align}
The measurement operator $ \operator{M}$ is defined by the detector's aperture function, \mbox{$w \in \Omega\times\mathbb{S}^2 $} (Fig.~\ref{fig:domain}b). The aperture function defines the manner in which the detector collects radiance, over a spatial and angular support. A given aperture and source pair $\left(w ,s\right)$ yields an element\begin{align}
 y_{w,s} = \operator{F}_{w,s} \left( \beta\right) = \operator{M}_w \operator{I}_s\left( \beta\right) = \langle w,I_s\rangle_{\Omega},
 \label{def:measmat}
\end{align}
for a specific extinction field $\beta$, where
\begin{align}
\langle \cdot \thinspace, \thinspace \cdot \rangle_{\Omega} \equiv \int_{\Omega}\int_{\mathbb{S}^2} \cdot \cdot d{\bm \omega} d{\bm x}.
 \label{def:innerprod}
\end{align}
For an idealized single-pixel detector positioned at ${\bm x}^\ast$, collecting radiation flowing in direction ${\bm \omega}^\ast$,
\begin{align}
 y_{w,s} = \left\langle \delta\left({\bm x}-{\bm x}^\ast\right)\delta\left({\bm \omega}-{\bm \omega}^\ast\right) \thinspace ,\thinspace I_s \right\rangle_{\Omega} = I_s\left({\bm x}^\ast,{\bm \omega}^\ast\right).
 \label{def:singlepixel}
\end{align}
We can define a matrix ${\bm Y}$ whose elements $y_{w,s}$ correspond to different source-detector configurations. A column of ${\bm Y}$ represents measurements by a single detector, for multiple sources. A row represents measurements by multiple detectors, for one particular light source. The vector ${\bm y}$ is the column stack of ${\bm Y}$.

\subsection{Optical tomography}
\label{subsec:tomography}

Using the operators defined in Sec.~\ref{subsec:opnotation} we express the process of optical tomography. Tomographic reconstruction is an estimator of $\beta$  that minimizes a defined cost
\begin{align}
 \hat{\beta} = \argmin_{\beta} \curly{\mathcal{E} \left[{\bm y}, \operator{F} \left(\beta\right) \right] + \alpha\Psi\left( \beta\right)},
 \label{eq:cost}
\end{align}
where $ \mathcal{E} \left[{\bm y}, \operator{F} \left(\beta\right) \right]$ is the {\em data fit} (fidelity) functional and $\Psi\left(\beta\right)$ is a {\em regularization} on the optical parameters. Here $\alpha$ is a tunable parameter, chosen in accordance with the noise level, to balance the two terms. \\

\begin{figure}[t]
    \centering
    \def\svgwidth{\columnwidth}
    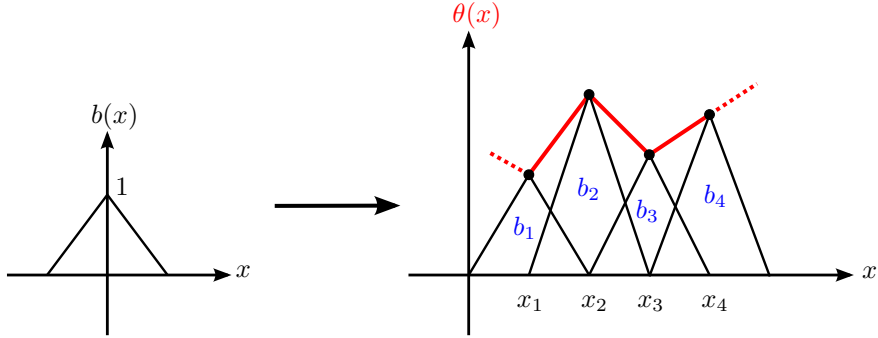
    \caption{Illustration of a 1D linear interpolation kernel}
    \label{fig:interpolation}
\end{figure}

\par Except for a few special geometric configurations \cite{davis2002cloud,arridge2009optical}, 3D tomographic recovery cannot be done analytically. Thus, the continuous function is often discretized (Fig.~\ref{fig:interpolation})
\begin{align}
 \beta \left({\bm x}\right)  = \sum_{k=1}^{N_{\rm grid}} \beta_k b_k\left({\bm x}\right),
 \label{eq:discrete}
\end{align}
where $\curly{\beta_k}_{k=1}^{N_{\rm grid}}$ are discrete parameters, $b_k\left({\bm x}\right)$ is an {\em interpolation kernel} and $N_{\rm grid}$ is the number of grid points used for discretization. We apply (\ref{eq:discrete}) to Eq.~(\ref{eq:cost}) to seek an estimator for \mbox{${\bm \beta} = \left(\beta_1,...,\beta_{N_{{\rm grid}}}\right)^T$}, where $(\thinspace\cdot\thinspace)^T$ denotes transposition.\\

\par We focus here on gradient-based optimization methods. The forward operator $\operator{F}$ is a non-linear function of the optical parameters. Thus, one approach \cite{arridge2009optical} estimates (\ref{eq:cost}) by linearizing $\operator{F}$. Suppose the solution is $\beta^{\delta} = \beta_0 + \delta \beta$, which is a perturbation of an initial guess $\beta_0$. It is possible to linearize $\operator{F}$ and solve $\delta \beta$ using a linear set for equations. However, this approach requires an initial guess very close to the true solution. Another approach \cite{abdoulaev2003three,gkioulekas2013inverse,klose1999iterative,park2000solution} iteratively estimates the gradient with respect to ${\bm \beta}$. However, this approach has high computational complexity, as it requires $\cal{O}\left(N_{{\rm grid}}\right)$ simulations of the forward model per iteration (i.e., for a single computation of the gradient). Forward model simulation is time consuming particularly in the presence of multiple scattering. Simulating the forward model per element of the gradient does not scale well as $N_{{\rm grid}}$ increases.\\

\par Our approach is also iterative, however each iteration is computationally simple, having run-time independent of $N_{\rm grid}$. The approach does not directly optimize the nonlinear $\operator{F}$. Instead, $\operator{I}$ is decomposed into a product of two operators. Each is optimized in time. This concept is analogous to {\em expectation minimization} (EM) optimization, which is used in DOT \cite{wang2010like,wang2011total,cao2007image}.

\section{Optimization Approach}
\label{sec:EM}

\subsection{Forward map decomposition}
\label{subsec:decomp}

In Sec.~\ref{sec:theor-backgr} the forward operator was defined in terms of the {\em radiance} field $I$. We now show that defining $\operator{F}$ in terms of the {\em in-scatter} field $J$ considerably speeds up the computation time. We decompose the radiance forward mapping into two operators, \mbox{$\operator{I} = \operator{T}\operator{J}$}, which we now introduce.

\begin{figure}[t]
    \centering
    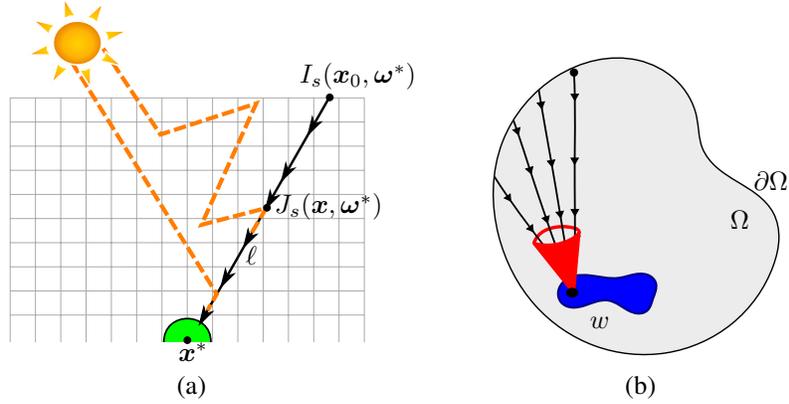
    \caption{(a) A line integral over the in-scatter field $J_s$.  (b) Integration over the field $f$ for an arbitrary aperture spatial and angular support $w$}
    \label{fig:lineint}
\end{figure}

For a particular radiation source $s \in \mathcal{S}$, let \mbox{$J_s = \operator{J}_s \left(  \beta \right)$} be the in-scatter field that satisfies Eq.~(\ref{eq:srcfun}), while $I_s$ satisfies Eqs. (\ref{eq:BC},\ref{eq:RTE}). Define $\mathcal{V}_s$ as the set of possible in-scatter fields for a particular source. The {\em in-scatter forward mapping}, $\operator{J}: \Theta \rightarrow \mathcal{V}$, is a mapping from the optical parameter domain to the in-scatter field range $\mathcal{V} = \bigcup_s \mathcal{V}_s$. We define the following transformation operator
\begin{align}
 \operator{T}: \mathcal{V} \rightarrow \mathcal{Z}, \qquad I_s = \operator{T}\left(\beta\right)J_s.
 \label{def:trans}
\end{align}
Eq.~(\ref{eq:srcfun}) defines a relation between a given light field $I$ and a corresponding in-scatter field $J$. The transformation operator $\operator{T}$, defined by Eq.~(\ref{def:Tdef}), defines the {\em inverse} relation: for a given in-scatter field $J$, a corresponding light field $I$ is attained. Consequently, we define the forward operator in terms of the in-scatter forward mapping
\begin{align}
 \operator{F} = \operator{M} \operator{T}\left(\beta\right) \operator{J}\left(\beta\right) : \Theta \rightarrow \mathcal{Y}.
 \label{def:fwdscat}
\end{align} \\
Define $\ell$ as a line of sight from ${\bm x}^\ast$ in direction $-{\bm \omega}^\ast$ (Fig.~\ref{fig:lineint}a). For a convex spatial domain, the intersection point between line of sight and the domain's boundary is unique and denoted by ${\bm x}_0 = \partial \Omega \cap \ell$  (Fig.~\ref{fig:lineint}a).

Operator $\operator{T}$ is defined by Eq.~(\ref{def:Tdef}). For source $s\in\mathcal{S}$, the response of an idealized single pixel with aperture \mbox{$w = \delta \left({\bm x} - {\bm x}^\ast\right)\delta\left({\bm \omega} - {\bm \omega}^\ast\right)$} is
\begin{align}
 y_{w,s} &= \operator{M}_w \operator{T} J_s  \nonumber \\
&=  I_s({\bm x}_0,{\bm \omega}^\ast)\exp{\left[-\int_{{\bm x}^\ast}^{{\bm x}_0}\beta\left({\bm r}\right)d{\bm r}\right]} + \int_{{\bm x}^\ast}^{{\bm x}_0} J_s({\bm x},{\bm \omega}^\ast) \beta\left({\bm x}\right)\exp{\left[-\int_{{\bm x}^\ast}^{{\bm x}}\beta\left({\bm r}\right)d{\bm r}\right]}d{\bm x}
\label{def:Jsinglepixel}
\end{align}
Eq.~(\ref{def:Jsinglepixel}) is simply an accumulation of the scattered radiance along the line of sight weighted by its corresponding attenuation factor (Fig.~\ref{fig:lineint}a). Applying a general measurement operation is a manner of integrating over the aperture function of the detector (Fig.~\ref{fig:lineint}b).

\begin{figure}[t]
    \centering
    \def\svgwidth{\columnwidth}
    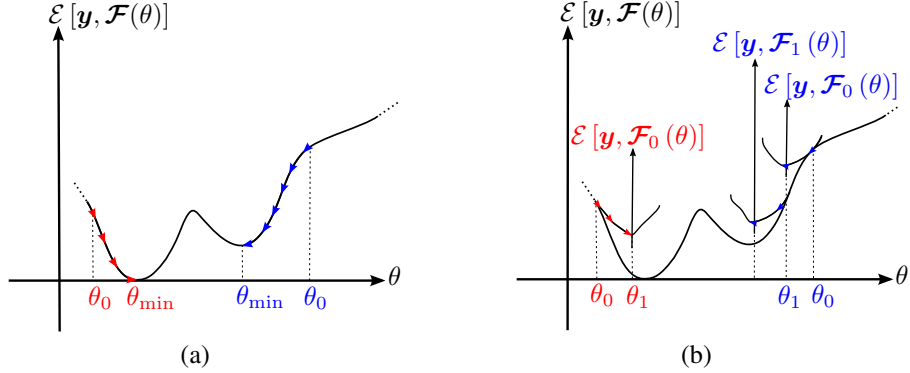
    \caption{(a) Gradient decent optimization. (b) Surrogate function iterative optimization. Both depend on the initial guess $\beta_0$}
    \label{fig:optimization}
\end{figure}

\subsection{Iterative surrogate function optimization}
Optimizing Eq.~(\ref{eq:cost}) directly over $\beta$ is computationally expensive. Finding the gradient direction in each iteration requires $\cal{O}\left(N_{\rm grid}\right)$ numerical computations of the forward model. With our approach, however, we iteratively keep $J_s$ constant and estimate ${\bm \beta}$ solely based on $\operator{T}$. Instead of optimizing through a function that is difficult to compute (Fig.~\ref{fig:optimization}b) we optimize ${\bm \beta}$ using a {\em surrogate function} \cite{lange2000optimization}, which is efficiently computed (Fig.~\ref{fig:optimization}a). We define the following iterative optimization process. Define $\beta_n$ as an estimate of $\beta$ in the ``$n$'' iteration. The first step consists of computing the in-scatter field, which corresponds to the current estimate $\beta_n$
$$J_n = \operator{J}\left(\beta_n\right).$$
In the second step, keeping the in-scatter field constant, we solve the following optimization problem to find our next estimate of $\beta$
\begin{align}
&\beta_{n+1} = \argmin_{ \beta} \curly{ \mathcal{E} \left[{\bm y}, \operator{F}_n \left(\beta\right) \right] + \alpha\Psi\left( \beta\right)}, \text{ with} \\
&\mathcal{E} \left[{\bm y}, \operator{F}_n \left(\beta\right) \right] = \left[{\bm y}-\operator{M}\operator{T}\left(\beta\right)J_n\right]^T {\bm \Sigma}_{\rm meas}^{-1} \left[{\bm y}-\operator{M}\operator{T}\left(\beta\right)J_n\right], \nonumber
\label{def:mstep}
\end{align}
where
\begin{align}
\operator{F}_n \left(\beta\right) = \operator{M}\operator{T}\left(\beta\right)J_n,
\end{align}
is the $n^{\rm th}$ surrogate function, and ${\bm \Sigma}_{\rm meas}$ is the covariance matrix of our measurements.  In the remainder, we will not make use of regularization (set $\alpha$ = 0).\\

\noindent Eq.~(\ref{def:mstep}) defines an iterative optimization process,\\
\begin{tabular}{ l l}
  {\tt i.}  & Start with an initial guess $\beta_0$.\\
  {\tt ii.} & Based on the current estimate $\beta_n$, numerically find the in-scatter field $J_n = \operator{J}\left(\beta_n \right)$.\\
  {\tt iii.} & Optimize (\ref{def:mstep}) to find the next estimate $\beta_{n+1}$.\\
  {\tt iv.} & Repeat steps {\tt ii-iii} until convergence.\\
\end{tabular}

\subsection{Scalability}
\label{subsec:scalability}

\begin{figure}[t]
    \centering
    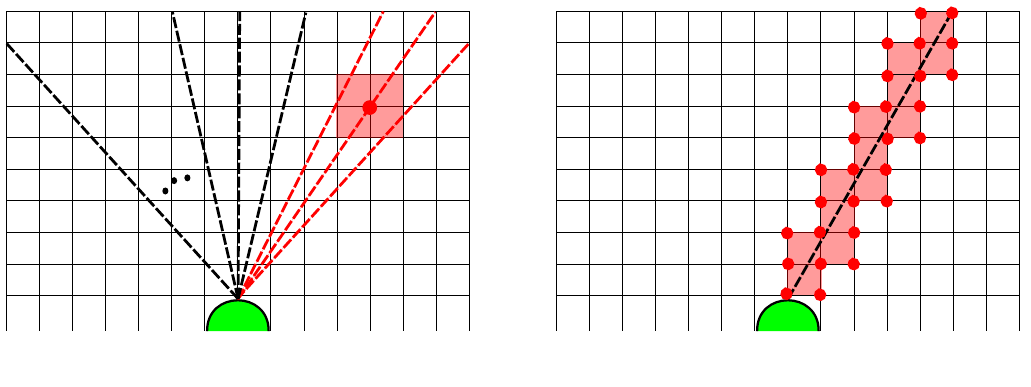
    \caption{(a) Each gradient element is computed by tracing all the rays. Only the red rays contribute to the computation of the red grid point gradient.  (b) A single trace of each ray is performed. While tracing each ray we compute the gradient contribution to each relevant grid point}
    \label{fig:scale}
\end{figure}

For a given measurement vector ${\bm y} \in \mathcal{Y}$, we solve the optimization of (\ref{def:mstep}) with a gradient-based method. We use the discretization described in (\ref{eq:discrete}). Assuming uncorrelated measurements
\begin{align}
{\bm \Sigma}_{\rm meas} = \diag{({\bm \sigma}^2_{\rm meas})} , \qquad {\bm \sigma}^2_{\rm meas} = \left(\sigma_1^2,...,\sigma_{N_{\rm meas}}^2 \right)^T,
\label{eq:nocorr}
\end{align}
where $N_{\rm meas} = N_w \times N_s$ is the number of measurements, $ N_w$ being the number of detectors and $N_s$ the number of sources.  In this study based on radiometry, we use $\sigma_i = 0.03 \times y_i$ (a 3\% noise due mostly to calibration error).

The task is thus to recover the gridded extinction $\curly{\beta_k}_{k=1}^{N_{\rm grid}}$. For this purpose we find the $k$'th gradient element. Without loss of generality we formulate the problem in terms of a single source and many detectors.
\begin{align}
\frac{\partial}{\partial \beta_k} \mathcal{E} \left[{\bm y}, \operator{F}_n \left({\bm \beta}\right) \right]
 = \sum_{w=1}^{N_{\rm meas}} \frac{1}{\sigma_{w}^2} \left[\operator{F}_{w,n}\left({\bm \beta}\right) - y_w\right]\operator{M}_{w}\left[\frac{\partial}{\partial \beta_k} \operator{T}\left({\bm \beta}\right)\right]J_n,
\label{eq:kgrad}
\end{align}
For an ideal single-pixel detector positioned at ${\bm x}^\ast$, collecting radiation flowing in direction ${\bm \omega}^\ast$ we get
\begin{align}
\operator{M}_{w}\left[\frac{\partial}{\partial \beta_k} \operator{T}\left({\bm \beta}\right)\right]J_n = A_{w,k} + B_{w,k}
\end{align}
where
\begin{align}
A_{w,k} = \left[-\int_{{\bm x}^\ast}^{{\bm x}_0}b_k\left({\bm r}\right)d{\bm r}\right]I({\bm x}_0,{\bm \omega}^\ast)\exp{\left[-\int_{{\bm x}^\ast}^{{\bm x}_0}\beta\left({\bm r}\right)d{\bm r}\right]} \end{align}
and
\begin{align}
B_{w,k} = \int_{{\bm x}^\ast}^{{\bm x}_0} J_n\left({\bm x},{\bm \omega}^\ast\right)\left[b_k({\bm x})-\beta({\bm x})\int_{{\bm x}^\ast}^{{\bm x}}b_k\left({\bm r}\right)d{\bm r}\right] \exp{\left[-\int_{{\bm x}^\ast}^{{\bm x}}\beta\left({\bm r}\right)d{\bm r}\right]}d{\bm x}.
\end{align}
Term $A$ results from the boundary illumination and can be readily computed. Term $B$ defines a line integral over a field, computed using back-projection of rays from the detector through the medium. A straightforward approach calculates each element of the gradient vector by summing all the back-projected rays from all the detectors (Alg.~\ref{Alg:pseudo}a). The complexity of this approach is $\cal{O}\left(N_{\rm meas} \times N_{\rm grid}\right)$ back-projecting operations. However, most of back-projected rays do not contribute to $B$ since the interpolation kernel $b_k({\bm x})$ typically has a small support region (Fig.~\ref{fig:scale}a). Instead, for each point along a ray we define a support region of grid points, according to the support region of the interpolation kernel (Fig.~\ref{fig:scale}b).

A specific ray will contribute to the computation of gradient elements associated with the grid points in the support region. In the case of a linear interpolation kernel, a support region is composed of the eight corner grid points that make a grid cell. Since the support region is typically small, for each point along the ray we only need to look at a finite amount of neighboring grid points (Alg.~\ref{Alg:pseudo}b). The complexity of this approach is $\cal{O}\left(N_{\rm meas} \right)$ back-projecting operations. Hence, the computational complexity is independent of the number of grid points which makes this approach scalable.

\begin{table}
\begin{subalgorithm}{.5\textwidth}
\begin{algorithmic}[1]
        \For {$k = 1 \to N_{\rm grid}$}
           \State ${\rm Grad}_k = 0$
            \For {$w = 1 \to N_{\rm meas}$}
             \State BP to compute $A_{w,k}$, $B_{w,k}$
             \State ${\rm Err} = \frac{1}{\sigma_w^2} \left[\operator{F}_n\left({\bm \beta}\right) - y_w\right]$
             \State ${\rm Grad}_k = {\rm Grad}_k \thickspace +$ \par
             \hskip\algorithmicindent\hskip\algorithmicindent $+ \thickspace {\rm Err} \left(A_{w,k} + B_{w,k}\right)$
            \EndFor
        \EndFor
\end{algorithmic}
\caption{Straightforward approach} \label{algo1}
\end{subalgorithm}%
\begin{subalgorithm}{.5\textwidth}
\begin{algorithmic}[1]
           \State ${\rm Grad} = {\bm 0}$
            \For {$w = 1 \to N_{\rm meas}$}
             \State ${\rm Err} = \frac{1}{\sigma_w^2} \left[\operator{F}_n\left({\bm \beta}\right) - y_w\right]$
             \For {$k \in $ {Support region}}
                \State BP to compute $A_{w,k}$,$B_{w,k}$
                \State ${\rm Grad}_k = {\rm Grad}_k \thickspace +$ \par
                	\hskip\algorithmicindent\hskip\algorithmicindent $+ \thickspace {\rm Err} \left(A_{w,k} + B_{w,k}\right)$
             \EndFor
            \EndFor
\end{algorithmic}
\caption{Scalable approach}\label{algo2}
\end{subalgorithm}
\captionsetup{labelformat=alglabel}
\caption{Comparison of the two approaches to compute the gradient of the surrogate function. ${\rm Grad} = \left({\rm Grad}_1,...,{\rm Grad}_{N_{\rm grid}}\right)$}
\label{Alg:pseudo}
\end{table}

\section{Application to remote sensing}
\label{sec:remote}

\subsection{General setup}
\label{subsec:setup}
\begin{figure}[t]
    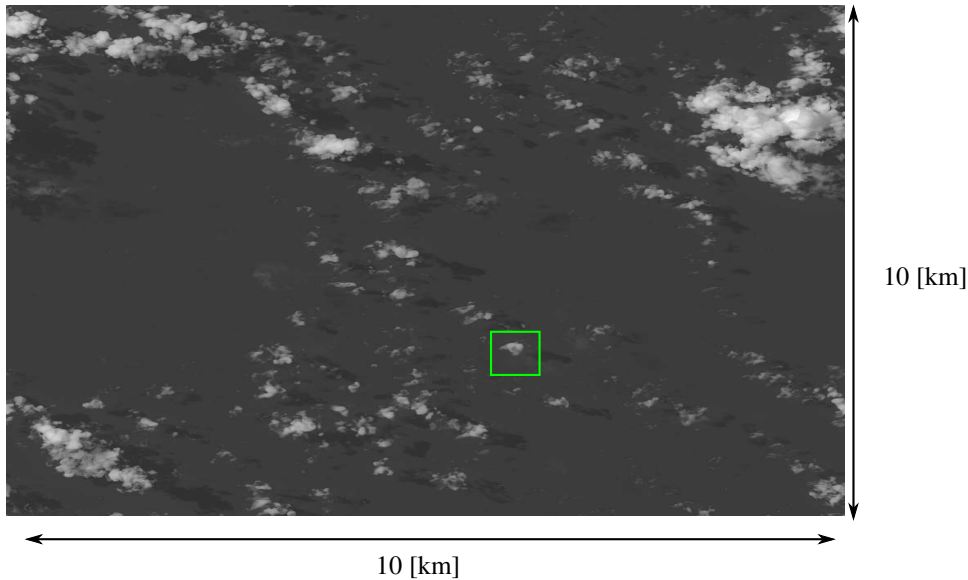
    \caption{Cumulus cloud field generated by the LES \cite{matheou2014large}. This nadir-viewing image was generated with the MYSTIC Monte Carlo code \cite{buras2011efficient,mayer2009radiative}. We selected the isolated cloud circled in green to test our tomographic reconstruction algorithm.}
    \label{fig:mystic_nadir}
\end{figure}

We use {\em large eddy simulation} (LES) \cite{chung2014large,matheou2014large} to generate the microphysical quantities of a realistic cloud field (Fig.~\ref{fig:mystic_nadir}). We use Mie scattering theory (implemented in the available code \cite{evans}) to convert the microphysical quantities (liquid water content from LES, assumed effective droplet radius $r_{\rm eff}$ and effective variance of the droplet size distribution $v_{\rm eff}$) into the required optical quantities (extinction coefficient, phase function, single scattering albedo). We take $r_{\rm eff} = 10 \mu$m and $v_{\rm eff} = 0.1$ for Gamma-distributed droplet sizes, and the wavelength is set to 672~nm. Finally, we add the extinction and phase function due to {\em Rayleigh} scattering by air molecules; aerosols are ignored in this simulation.

We seek to recover the extinction of the cloud droplets on a cartesian grid, where the air is taken to be a known parameter dependent only on the height. The boundary conditions for the domain are
  \begin{itemize}
  \item Collimated solar radiation at the top of the atmosphere (TOA), at a zenith angle of $60^\circ$.
  \item Open boundary for the side faces, to approximate an isolated cloud observed at a wavelength where the Rayleigh sky is almost black (Rayleigh optical thickness of the atmosphere is only 0.04532 at 672~nm).
  \item {\em Lambertian} reflectance at the surface (Earth) with an albedo of 0.05, to mimic the dark ocean surface.
\end{itemize}
At $\lambda$ = 672~nm, liquid water and air are non-absorbing, so the single scattering albedo $\varpi$ is everywhere unity. In order to find the in-scatter field $J_n$, we use SHDOM \cite{evans1998}.

\subsection{SHDOM}
\label{subsec:SHDOM}

Running the forward model is a balance between speed and accuracy. Monte Carlo methods can handle very complex optical media. However, they compute radiometric quantities by random sampling the infinite domain of possible light paths. This introduces stochastic noise at the output, which can be controlled by increasing the number of samples (photon paths) and variance reduction techniques. When many radiometric outputs of the same scene are sought, as in the case of many viewpoints, a model that solves the {\em RTE} directly has a much more favorable runtime \cite{pincus2009}. A discrete ordinates representation models the flow of radiant energy in the domain easily and intuitively  \cite{truelove1987discrete,sanchez1994,evans1998,evans2005}. The SHDOM model uses a grid for the spatial dependency, and a linear interpolation kernel $b_k({\bm x})$. {\em Spherical Harmonic} expansion \cite{kajiya1984,hancourant2008} is used to compute angular integrals. SHDOM solves the forward model 3D RTE in an integral form for the source function $J\xw$. It can adaptively truncate negligible coefficients in the series expansion. It also has adaptive spatial grid capability and multi-grid acceleration.

\subsection{Simulation Results}

\begin{figure}[t]
    \centering
    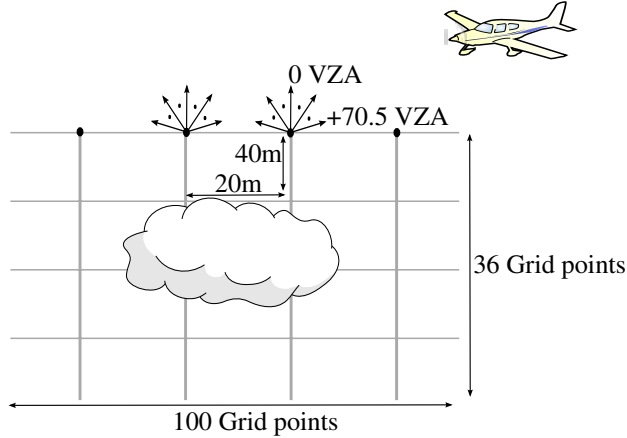
    \caption{AirMSPI measurements: 9 zenith viewing angles (VZAs) in the along track direction. There are also rows of 36 across-track pixels (20~m width).}
    \label{fig:airMSPI}
\end{figure}

We simulate an atmosphere of size $2\,{\rm km} \times 0.72\,{\rm km} \times 1.44\,{\rm km} $. The unknown extinction is composed of $100 \times 36 \times 36$ grid points (129,600 unknowns).  The measurements are taken to approximate the resolution of {\em AirMSPI} (Airborne Multi-angle Spectro-Polarimeter Imager) \cite{diner2013airborne}, which is an airborne sensor that can sample radiance at 10~m resolution and, to emulate the space-borne Multi-angle Imaging Spectro-Radiometer (MISR) \cite{diner1998multi}, at 9 viewing zenith angles: $\left\{\pm70.5^\circ,\pm60^\circ,\pm45.6^\circ,\pm26.1^\circ,0^\circ\right\}$ where $\pm$ indicates two azimuthal half-planes at $180^\circ$ apart (Fig.~\ref{fig:airMSPI}). We simulate the measurements using SHDOM and add 3\% gaussian noise to simulate the radiometric calibration noise, which is the dominant factor for this sensor. We initialize our algorithm with ${\bm \beta} = {\bm 0}$ (an atmosphere containing only air molecules).

\begin{figure}[H]
    \centering
    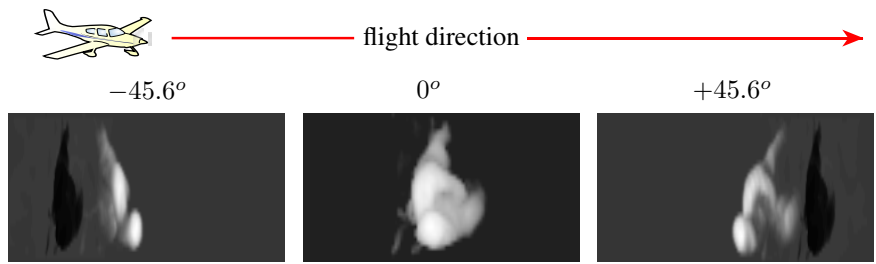
    \caption{A simulated fly-over of the AirMSPI using SHDOM to generate the radiance measurements at 3 of the 9 viewing angles. The cloud extinction field is that of the recovered cloud.}
    \label{fig:fly_over}
\end{figure}

\begin{figure}[H]
    \centering
    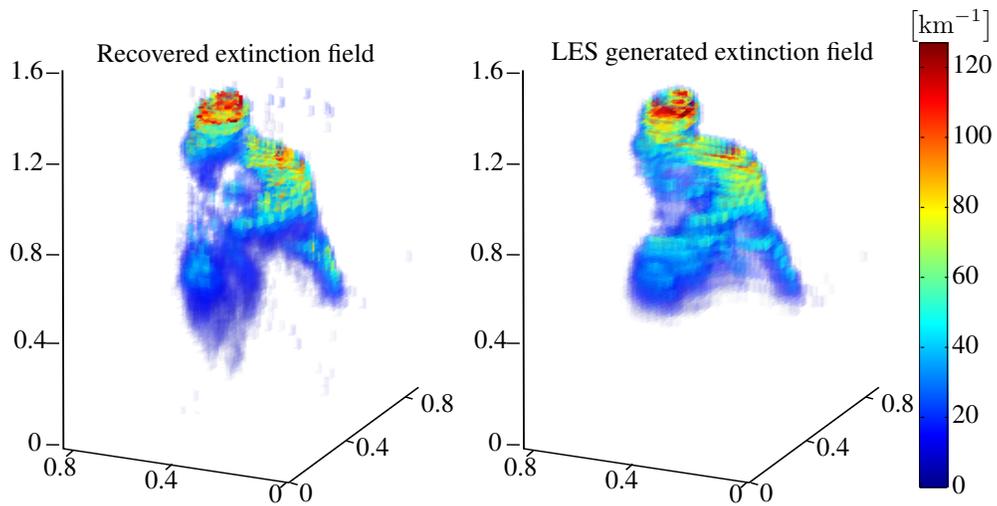
    \caption{A volumetric comparison between the ground truth LES generated cloud and the recovered cloud. Axes are in km.}
    \label{fig:recovery}
\end{figure}

Simulated images as viewed from the AirMSPI instrument flying over the recovered cloud are shown in Fig.~\ref{fig:fly_over}. The converged reconstruction is displayed in Fig.~\ref{fig:recovery} along with the ground-truth. We compare the total extinction recovered, relative to the total extinction, with the LES/Mie-based truth. We define a relative error measure
$$ \text{Absolute Relative Error} = \frac{\left|{\bm \beta}^{\rm ground\,truth} - {\bm \beta}^{\rm recovered}\right|}{\sum_k \beta_k^{\rm ground\,truth}}$$
The performance for the recovery is summarized in Table \ref{tab:results}.

The extinction coefficient summed over the grid, $\sum_k \beta_k$ (in km$^{-1}$), can be converted into a domain-average vertical optical thickness ($\times 0.04\,{\rm km} / (36 \times 100)$), yielding 0.956.  The recovered value is 1.089, a 14\% overestimate.  Most of this vertically-integrated opacity is of course in the cloud (the Rayleigh optical thickness of the lowest 1.44~km layer being only 0.0075), which covers $\sim$20\% of the horizontal domain (based on ${\rm LWC}>10^{-5} {\rm g} \cdot {\rm m}^{-3}$ threshold). 

\section{Summary and Outlook}

We derive a novel iterative optimization method to perform radiative transfer tomography. The unknown $\beta$ directly affects $\operator{T}(\beta)$. From Eq.~(\ref{def:Tdef}), this principle can only retrieve parameters that relate to the extinction (we cannot estimate the phase function using this formulation). Sec.~\ref{subsec:scalability} explains the computational advantage of this approach. We show an application to remote sensing of Earth's atmosphere (Sec.~\ref{sec:remote}), however, this approach could potentially be applied to preform optical tomography of biological tissues. We use SHDOM as our forward mapping engine. Nevertheless, it is possible to use different radiative transfer engines, such as {\em Monte Carlo}. Further enhancement of scalability may be obtained by use of adjoint operators \cite{martin2014adjoint}, a subject intended for further research. Another extension is to determine how well one can reconstruct an isolated cloud's shape and internal structure using satellite data from MISR (275~m pixels), which has global coverage, rather than AirMSPI (10 to 20~m pixels), which is deployed only at field campaigns.  We anticipate that optimization using this coarser data will require careful regularization.

\begin{table}
\centering
\begin{tabular}{l l}
\toprule[1.1pt]
    $\sum_k\beta^{\rm ground\,truth}_k$ & 8.606 10$^{+4}$ km$^{-1}$\\
    $\sum_k\beta^{\rm recovered}_k$     & 9.800 10$^{+4}$ km$^{-1}$\\
    Mean absolute relative error        & 4.672 10$^{-6}$\\
	Maximum absolute relative error     & 8.025 10$^{-4}$\\
\bottomrule[1.1pt]
\end {tabular}
\caption {Recovery results}
\label{tab:results}
\end{table}

\section*{Acknowledgments}
We are grateful to David J. Diner for fruitful discussions and his support at all levels. We also thank Anat Levin, Vadim Holodovsky, Michael Garay and Zheng Qu for useful discussions, and Johanan Erez, Ina Talmon and Dani Yagodin for technical support. We would like to thank Frank Evans for his SHDOM software package. Yoav Schechner is a Landau Fellow - supported by the Taub Foundation. His work is supported by the US-Israel Binational Science Foundation (BSF grant 2012202) and the E. and J. Bishop Research Fund. This work was conducted in the  Ollendorff Minerva Center. Minerva is funded through the BMBF. The research of Anthony Davis is performed at the Jet Propulsion Laboratory, California Institute of Technology, under contract with NASA and with support by NASA's Earth Science Technology Office Advanced Information Systems Technology Program.

\bibliographystyle{unsrt}
\bibliography{Fast_3d_RT_Tomography}

\end{document}

%% file: domain.pdf_tex
\begingroup%
  \makeatletter%
  \providecommand\color[2][]{%
    \errmessage{(Inkscape) Color is used for the text in Inkscape, but the package 'color.sty' is not loaded}%
    \renewcommand\color[2][]{}%
  }%
  \providecommand\transparent[1]{%
    \errmessage{(Inkscape) Transparency is used (non-zero) for the text in Inkscape, but the package 'transparent.sty' is not loaded}%
    \renewcommand\transparent[1]{}%
  }%
  \providecommand\rotatebox[2]{#2}%
  \ifx\svgwidth\undefined%
    \setlength{\unitlength}{278.23200684bp}%
    \ifx\svgscale\undefined%
      \relax%
    \else%
      \setlength{\unitlength}{\unitlength * \real{\svgscale}}%
    \fi%
  \else%
    \setlength{\unitlength}{\svgwidth}%
  \fi%
  \global\let\svgwidth\undefined%
  \global\let\svgscale\undefined%
  \makeatother%
  \begin{picture}(1,0.51656069)%
    \put(0,0){\includegraphics[width=\unitlength]{domain.pdf}}%
    \put(0.34071546,0.1935725){\color[rgb]{0,0,0}\makebox(0,0)[lb]{\smash{$\Omega$}}}%
    \put(0.34210093,0.26577748){\color[rgb]{0,0,0}\makebox(0,0)[lb]{\smash{$\partial \Omega$}}}%
    \put(0.19921565,0.49592257){\color[rgb]{0,0,0}\makebox(0,0)[lb]{\smash{$I_{\rm incident}$}}}%
    \put(0.12307383,0.2985001){\color[rgb]{0,0,0}\makebox(0,0)[lb]{\smash{$I \xw $}}}%
    \put(0.8052565,0.45739905){\makebox(0,0)[lb]{\smash{$s$}}}%
    \put(0.67385525,0.27718625){\color[rgb]{0,0,0}\makebox(0,0)[lb]{\smash{$I_s$}}}%
    \put(0.20213198,0.00614934){\color[rgb]{0,0,0}\makebox(0,0)[lb]{\smash{(a)}}}%
    \put(0.71570743,0.00658316){\color[rgb]{0,0,0}\makebox(0,0)[lb]{\smash{(b)}}}%
    \put(0.69770838,0.08633013){\color[rgb]{0,0,0}\makebox(0,0)[lb]{\smash{$w$}}}%
    \put(0.84317813,0.19643488){\color[rgb]{0,0,0}\makebox(0,0)[lb]{\smash{$\Omega$}}}%
    \put(0.8835917,0.23821522){\color[rgb]{0,0,0}\makebox(0,0)[lb]{\smash{$\partial \Omega$}}}%
    \put(0.01710886,0.2570344){\color[rgb]{0,0,0}\makebox(0,0)[lb]{\smash{${\bm \vartheta}$}}}%
  \end{picture}%
\endgroup%

%% file: interpolation.pdf_tex
\begingroup%
  \makeatletter%
  \providecommand\color[2][]{%
    \errmessage{(Inkscape) Color is used for the text in Inkscape, but the package 'color.sty' is not loaded}%
    \renewcommand\color[2][]{}%
  }%
  \providecommand\transparent[1]{%
    \errmessage{(Inkscape) Transparency is used (non-zero) for the text in Inkscape, but the package 'transparent.sty' is not loaded}%
    \renewcommand\transparent[1]{}%
  }%
  \providecommand\rotatebox[2]{#2}%
  \ifx\svgwidth\undefined%
    \setlength{\unitlength}{181.72734375bp}%
    \ifx\svgscale\undefined%
      \relax%
    \else%
      \setlength{\unitlength}{\unitlength * \real{\svgscale}}%
    \fi%
  \else%
    \setlength{\unitlength}{\svgwidth}%
  \fi%
  \global\let\svgwidth\undefined%
  \global\let\svgscale\undefined%
  \makeatother%
  \begin{picture}(1,0.37998633)%
    \put(0,0){\includegraphics[width=\unitlength]{interpolation.pdf}}%
    \put(0.09464729,0.23551767){\color[rgb]{0,0,0}\makebox(0,0)[lb]{\smash{$b(x)$}}}%
    \put(0.25312646,0.0660342){\color[rgb]{0,0,0}\makebox(0,0)[lb]{\smash{$x$}}}%
    \put(0.12106048,0.15627808){\color[rgb]{0,0,0}\makebox(0,0)[lb]{\smash{$1$}}}%
    \put(0.49084523,0.34557265){\color[rgb]{1,0,0}\makebox(0,0)[lb]{\smash{$\color{red}\theta(x)$}}}%
    \put(0.93986957,0.06515374){\color[rgb]{0,0,0}\makebox(0,0)[lb]{\smash{$x$}}}%
    \put(0.56128042,0.0330165){\color[rgb]{0,0,0}\makebox(0,0)[lb]{\smash{$x_1$}}}%
    \put(0.63171561,0.0330165){\color[rgb]{0,0,0}\makebox(0,0)[lb]{\smash{$x_2$}}}%
    \put(0.6933464,0.0330165){\color[rgb]{0,0,0}\makebox(0,0)[lb]{\smash{$x_3$}}}%
    \put(0.76378159,0.0330165){\color[rgb]{0,0,0}\makebox(0,0)[lb]{\smash{$x_4$}}}%
    \put(0.55729281,0.11178383){\color[rgb]{0,0,0}\makebox(0,0)[lb]{\smash{$\color{blue}b_1$}}}%
    \put(0.627525,0.15526517){\color[rgb]{0,0,0}\makebox(0,0)[lb]{\smash{$\color{blue}b_2$}}}%
    \put(0.69167744,0.13094622){\color[rgb]{0,0,0}\makebox(0,0)[lb]{\smash{$\color{blue}b_3$}}}%
    \put(0.76614296,0.14675354){\color[rgb]{0,0,0}\makebox(0,0)[lb]{\smash{$\color{blue}b_4$}}}%
  \end{picture}%
\endgroup%

%% file: line_integral.pdf_tex
\begingroup%
  \makeatletter%
  \providecommand\color[2][]{%
    \errmessage{(Inkscape) Color is used for the text in Inkscape, but the package 'color.sty' is not loaded}%
    \renewcommand\color[2][]{}%
  }%
  \providecommand\transparent[1]{%
    \errmessage{(Inkscape) Transparency is used (non-zero) for the text in Inkscape, but the package 'transparent.sty' is not loaded}%
    \renewcommand\transparent[1]{}%
  }%
  \providecommand\rotatebox[2]{#2}%
  \ifx\svgwidth\undefined%
    \setlength{\unitlength}{297.06398926bp}%
    \ifx\svgscale\undefined%
      \relax%
    \else%
      \setlength{\unitlength}{\unitlength * \real{\svgscale}}%
    \fi%
  \else%
    \setlength{\unitlength}{\svgwidth}%
  \fi%
  \global\let\svgwidth\undefined%
  \global\let\svgscale\undefined%
  \makeatother%
  \begin{picture}(1,0.5041338)%
    \put(0,0){\includegraphics[width=\unitlength]{line_integral.pdf}}%
    \put(0.21397036,0.04611401){\color[rgb]{0,0,0}\makebox(0,0)[lb]{\smash{${\bm x}^\ast$}}}%
    \put(0.29930091,0.16833002){\color[rgb]{0,0,0}\makebox(0,0)[lb]{\smash{$\ell$}}}%
    \put(0.36756538,0.39872249){\color[rgb]{0,0,0}\makebox(0,0)[lb]{\smash{$I_s({\bm x}_0,{\bm \omega}^\ast)$}}}%
    \put(0.73590883,0.087266){\color[rgb]{0,0,0}\makebox(0,0)[lb]{\smash{$w$}}}%
    \put(0.91368079,0.21383965){\color[rgb]{0,0,0}\makebox(0,0)[lb]{\smash{$\Omega$}}}%
    \put(0.94354644,0.26219363){\color[rgb]{0,0,0}\makebox(0,0)[lb]{\smash{$\partial \Omega$}}}%
    \put(0.3362775,0.23344743){\color[rgb]{0,0,0}\makebox(0,0)[lb]{\smash{$J_s({\bm x},{\bm \omega}^\ast)$}}}%
    \put(0.21183225,0.00453305){\color[rgb]{0,0,0}\makebox(0,0)[lb]{\smash{(a)}}}%
    \put(0.78059415,0.00453305){\color[rgb]{0,0,0}\makebox(0,0)[lb]{\smash{(b)}}}%
  \end{picture}%
\endgroup%

%% file: optimization.pdf_tex
\begingroup%
  \makeatletter%
  \providecommand\color[2][]{%
    \errmessage{(Inkscape) Color is used for the text in Inkscape, but the package 'color.sty' is not loaded}%
    \renewcommand\color[2][]{}%
  }%
  \providecommand\transparent[1]{%
    \errmessage{(Inkscape) Transparency is used (non-zero) for the text in Inkscape, but the package 'transparent.sty' is not loaded}%
    \renewcommand\transparent[1]{}%
  }%
  \providecommand\rotatebox[2]{#2}%
  \ifx\svgwidth\undefined%
    \setlength{\unitlength}{214.8501709bp}%
    \ifx\svgscale\undefined%
      \relax%
    \else%
      \setlength{\unitlength}{\unitlength * \real{\svgscale}}%
    \fi%
  \else%
    \setlength{\unitlength}{\svgwidth}%
  \fi%
  \global\let\svgwidth\undefined%
  \global\let\svgscale\undefined%
  \makeatother%
  \begin{picture}(1,0.41170113)%
    \put(0,0){\includegraphics[width=\unitlength]{optimization.pdf}}%
    \put(0.04671215,0.38042207){\color[rgb]{0,0,0}\makebox(0,0)[lb]{\smash{$\mathcal{E}\left[{\bm y},\operator{F}(\theta)\right]$}}}%
    \put(0.41887555,0.09341782){\color[rgb]{0,0,0}\makebox(0,0)[lb]{\smash{$\theta$}}}%
    \put(0.74030952,0.00783612){\color[rgb]{0,0,0}\makebox(0,0)[lb]{\smash{(b)}}}%
    \put(0.19150411,0.00783612){\color[rgb]{0,0,0}\makebox(0,0)[lb]{\smash{(a)}}}%
    \put(0.60752513,0.38260382){\color[rgb]{0,0,0}\makebox(0,0)[lb]{\smash{$\mathcal{E}\left[{\bm y},\operator{F}(\theta)\right]$}}}%
    \put(0.97532501,0.09450871){\color[rgb]{0,0,0}\makebox(0,0)[lb]{\smash{$\theta$}}}%
    \put(0.83135771,0.30356546){\color[rgb]{0,0,0}\makebox(0,0)[lb]{\smash{$\color{blue}\mathcal{E}\left[{\bm y},\operator{F}_0\left(\theta\right)\right]$}}}%
    \put(0.77466122,0.35065496){\color[rgb]{0,0,0}\makebox(0,0)[lb]{\smash{$\color{blue}\mathcal{E}\left[{\bm y},\operator{F}_1\left(\theta\right)\right]$}}}%
    \put(0.61919087,0.24840535){\color[rgb]{0,0,0}\makebox(0,0)[lb]{\smash{$\color{red} \mathcal{E}\left[{\bm y},\operator{F}_0\left(\theta\right)\right]$}}}%
    \put(0.3266019,0.07177603){\color[rgb]{0,0,0}\makebox(0,0)[lb]{\smash{$\color{blue}\theta_0$}}}%
    \put(0.25236412,0.07261236){\color[rgb]{0,0,0}\makebox(0,0)[lb]{\smash{$\color{blue}\theta_{\rm min}$}}}%
    \put(0.09028897,0.07188511){\color[rgb]{0,0,0}\makebox(0,0)[lb]{\smash{$\color{red}\theta_0$}}}%
    \put(0.13233135,0.07188511){\color[rgb]{0,0,0}\makebox(0,0)[lb]{\smash{$\color{red}\theta_{\rm min}$}}}%
    \put(0.84616454,0.07304144){\color[rgb]{0,0,0}\makebox(0,0)[lb]{\smash{$\color{blue}\theta_1$}}}%
    \put(0.88324548,0.07323053){\color[rgb]{0,0,0}\makebox(0,0)[lb]{\smash{$\color{blue}\theta_0$}}}%
    \put(0.64062692,0.0743214){\color[rgb]{0,0,0}\makebox(0,0)[lb]{\smash{$\color{red}\theta_0$}}}%
    \put(0.68004392,0.07304144){\color[rgb]{0,0,0}\makebox(0,0)[lb]{\smash{$\color{red}\theta_1$}}}%
  \end{picture}%
\endgroup%

%% file: scalability.pdf_tex
\begingroup%
  \makeatletter%
  \providecommand\color[2][]{%
    \errmessage{(Inkscape) Color is used for the text in Inkscape, but the package 'color.sty' is not loaded}%
    \renewcommand\color[2][]{}%
  }%
  \providecommand\transparent[1]{%
    \errmessage{(Inkscape) Transparency is used (non-zero) for the text in Inkscape, but the package 'transparent.sty' is not loaded}%
    \renewcommand\transparent[1]{}%
  }%
  \providecommand\rotatebox[2]{#2}%
  \ifx\svgwidth\undefined%
    \setlength{\unitlength}{296.76799316bp}%
    \ifx\svgscale\undefined%
      \relax%
    \else%
      \setlength{\unitlength}{\unitlength * \real{\svgscale}}%
    \fi%
  \else%
    \setlength{\unitlength}{\svgwidth}%
  \fi%
  \global\let\svgwidth\undefined%
  \global\let\svgscale\undefined%
  \makeatother%
  \begin{picture}(1,0.37486522)%
    \put(0,0){\includegraphics[width=\unitlength]{scalability.pdf}}%
    \put(0.21046389,0.00576489){\color[rgb]{0,0,0}\makebox(0,0)[lb]{\smash{(a)}}}%
    \put(0.74421417,0.00576489){\color[rgb]{0,0,0}\makebox(0,0)[lb]{\smash{(b)}}}%
  \end{picture}%
\endgroup%

%% file: mystic_nadir.pdf_tex
\begingroup%
  \makeatletter%
  \providecommand\color[2][]{%
    \errmessage{(Inkscape) Color is used for the text in Inkscape, but the package 'color.sty' is not loaded}%
    \renewcommand\color[2][]{}%
  }%
  \providecommand\transparent[1]{%
    \errmessage{(Inkscape) Transparency is used (non-zero) for the text in Inkscape, but the package 'transparent.sty' is not loaded}%
    \renewcommand\transparent[1]{}%
  }%
  \providecommand\rotatebox[2]{#2}%
  \ifx\svgwidth\undefined%
    \setlength{\unitlength}{384bp}%
    \ifx\svgscale\undefined%
      \relax%
    \else%
      \setlength{\unitlength}{\unitlength * \real{\svgscale}}%
    \fi%
  \else%
    \setlength{\unitlength}{\svgwidth}%
  \fi%
  \global\let\svgwidth\undefined%
  \global\let\svgscale\undefined%
  \makeatother%
  \begin{picture}(1,0.56687978)%
    \put(0,0){\includegraphics[width=\unitlength]{mystic_nadir.pdf}}%
    \put(0.86170261,0.29109053){\color[rgb]{0,0,0}\makebox(0,0)[lb]{\smash{10 [km]}}}%
    \put(0.36411562,0.00657997){\color[rgb]{0,0,0}\makebox(0,0)[lb]{\smash{10 [km]}}}%
  \end{picture}%
\endgroup%

%% file: airMSPI.pdf_tex
\begingroup%
  \makeatletter%
  \providecommand\color[2][]{%
    \errmessage{(Inkscape) Color is used for the text in Inkscape, but the package 'color.sty' is not loaded}%
    \renewcommand\color[2][]{}%
  }%
  \providecommand\transparent[1]{%
    \errmessage{(Inkscape) Transparency is used (non-zero) for the text in Inkscape, but the package 'transparent.sty' is not loaded}%
    \renewcommand\transparent[1]{}%
  }%
  \providecommand\rotatebox[2]{#2}%
  \ifx\svgwidth\undefined%
    \setlength{\unitlength}{248bp}%
    \ifx\svgscale\undefined%
      \relax%
    \else%
      \setlength{\unitlength}{\unitlength * \real{\svgscale}}%
    \fi%
  \else%
    \setlength{\unitlength}{\svgwidth}%
  \fi%
  \global\let\svgwidth\undefined%
  \global\let\svgscale\undefined%
  \makeatother%
  \begin{picture}(1,0.67783873)%
    \put(0,0){\includegraphics[width=\unitlength]{airMSPI.pdf}}%
    \put(0.60454658,0.47975939){\color[rgb]{0,0,0}\makebox(0,0)[lb]{\smash{+70.5 VZA}}}%
    \put(0.54621185,0.54368963){\color[rgb]{0,0,0}\makebox(0,0)[lb]{\smash{0 VZA}}}%
    \put(0.43464191,0.37863024){\color[rgb]{0,0,0}\makebox(0,0)[lb]{\smash{20m}}}%
    \put(0.4652871,0.42298508){\color[rgb]{0,0,0}\makebox(0,0)[lb]{\smash{40m}}}%
    \put(0.37012578,0.01653347){\color[rgb]{0,0,0}\makebox(0,0)[lb]{\smash{100 Grid points}}}%
    \put(0.82738384,0.25443666){\color[rgb]{0,0,0}\makebox(0,0)[lb]{\smash{36 Grid points}}}%
  \end{picture}%
\endgroup%

%% file: fly_over.pdf_tex
\begingroup%
  \makeatletter%
  \providecommand\color[2][]{%
    \errmessage{(Inkscape) Color is used for the text in Inkscape, but the package 'color.sty' is not loaded}%
    \renewcommand\color[2][]{}%
  }%
  \providecommand\transparent[1]{%
    \errmessage{(Inkscape) Transparency is used (non-zero) for the text in Inkscape, but the package 'transparent.sty' is not loaded}%
    \renewcommand\transparent[1]{}%
  }%
  \providecommand\rotatebox[2]{#2}%
  \ifx\svgwidth\undefined%
    \setlength{\unitlength}{325.83317871bp}%
    \ifx\svgscale\undefined%
      \relax%
    \else%
      \setlength{\unitlength}{\unitlength * \real{\svgscale}}%
    \fi%
  \else%
    \setlength{\unitlength}{\svgwidth}%
  \fi%
  \global\let\svgwidth\undefined%
  \global\let\svgscale\undefined%
  \makeatother%
  \begin{picture}(1,0.2925536)%
    \put(0,0){\includegraphics[width=\unitlength]{fly_over.pdf}}%
    \put(0.40993523,0.25122184){\color[rgb]{0,0,0}\makebox(0,0)[lb]{\smash{flight direction}}}%
    \put(0.79135345,0.18904514){\color[rgb]{0,0,0}\makebox(0,0)[lb]{\smash{$+45.6^o$ }}}%
    \put(0.47174142,0.18904514){\color[rgb]{0,0,0}\makebox(0,0)[lb]{\smash{$0^o$ }}}%
    \put(0.1147287,0.18816826){\color[rgb]{0,0,0}\makebox(0,0)[lb]{\smash{$-45.6^o$ }}}%
  \end{picture}%
\endgroup%

%% file: recovery_volume.pdf_tex
\begingroup%
  \makeatletter%
  \providecommand\color[2][]{%
    \errmessage{(Inkscape) Color is used for the text in Inkscape, but the package 'color.sty' is not loaded}%
    \renewcommand\color[2][]{}%
  }%
  \providecommand\transparent[1]{%
    \errmessage{(Inkscape) Transparency is used (non-zero) for the text in Inkscape, but the package 'transparent.sty' is not loaded}%
    \renewcommand\transparent[1]{}%
  }%
  \providecommand\rotatebox[2]{#2}%
  \ifx\svgwidth\undefined%
    \setlength{\unitlength}{349.43999023bp}%
    \ifx\svgscale\undefined%
      \relax%
    \else%
      \setlength{\unitlength}{\unitlength * \real{\svgscale}}%
    \fi%
  \else%
    \setlength{\unitlength}{\svgwidth}%
  \fi%
  \global\let\svgwidth\undefined%
  \global\let\svgscale\undefined%
  \makeatother%
  \begin{picture}(1,0.54136907)%
    \put(0,0){\includegraphics[width=\unitlength]{recovery_volume.pdf}}%
    \put(0.46430969,0.47988917){\color[rgb]{0,0,0}\makebox(0,0)[lb]{\smash{1.6}}}%
    \put(0.46447205,0.38155998){\color[rgb]{0,0,0}\makebox(0,0)[lb]{\smash{1.2}}}%
    \put(0.4642633,0.28380544){\color[rgb]{0,0,0}\makebox(0,0)[lb]{\smash{0.8}}}%
    \put(0.46841898,0.18991771){\color[rgb]{0,0,0}\makebox(0,0)[lb]{\smash{0.4}}}%
    \put(0.48366491,0.07922964){\color[rgb]{0,0,0}\makebox(0,0)[lb]{\smash{0}}}%
    \put(0.74358896,0.02310614){\color[rgb]{0,0,0}\makebox(0,0)[lb]{\smash{0}}}%
    \put(0.60962809,0.03869602){\color[rgb]{0,0,0}\makebox(0,0)[lb]{\smash{0.4}}}%
    \put(0.50065996,0.05272689){\color[rgb]{0,0,0}\makebox(0,0)[lb]{\smash{0.8}}}%
    \put(0.77645983,0.02420853){\color[rgb]{0,0,0}\makebox(0,0)[lb]{\smash{0}}}%
    \put(0.83756151,0.0737732){\color[rgb]{0,0,0}\makebox(0,0)[lb]{\smash{0.4}}}%
    \put(0.90754107,0.1213223){\color[rgb]{0,0,0}\makebox(0,0)[lb]{\smash{0.8}}}%
    \put(0.06130275,0.4990769){\color[rgb]{0,0,0}\makebox(0,0)[lb]{\smash{Recovered extinction field}}}%
    \put(0.55178421,0.5011648){\color[rgb]{0,0,0}\makebox(0,0)[lb]{\smash{LES generated extinction field}}}%
    \put(-0.03216999,0.47988917){\color[rgb]{0,0,0}\makebox(0,0)[lb]{\smash{1.6}}}%
    \put(-0.03200763,0.38155998){\color[rgb]{0,0,0}\makebox(0,0)[lb]{\smash{1.2}}}%
    \put(-0.03221638,0.28380544){\color[rgb]{0,0,0}\makebox(0,0)[lb]{\smash{0.8}}}%
    \put(-0.0280607,0.18991771){\color[rgb]{0,0,0}\makebox(0,0)[lb]{\smash{0.4}}}%
    \put(-0.01281477,0.07922964){\color[rgb]{0,0,0}\makebox(0,0)[lb]{\smash{0}}}%
    \put(0.24710928,0.02310614){\color[rgb]{0,0,0}\makebox(0,0)[lb]{\smash{0}}}%
    \put(0.11314841,0.03869602){\color[rgb]{0,0,0}\makebox(0,0)[lb]{\smash{0.4}}}%
    \put(0.00418028,0.05272689){\color[rgb]{0,0,0}\makebox(0,0)[lb]{\smash{0.8}}}%
    \put(0.27998015,0.02420853){\color[rgb]{0,0,0}\makebox(0,0)[lb]{\smash{0}}}%
    \put(0.34108183,0.0737732){\color[rgb]{0,0,0}\makebox(0,0)[lb]{\smash{0.4}}}%
    \put(0.41106139,0.1213223){\color[rgb]{0,0,0}\makebox(0,0)[lb]{\smash{0.8}}}%
    \put(0.9838206,0.03317238){\color[rgb]{0,0,0}\makebox(0,0)[lb]{\smash{0}}}%
    \put(0.9838206,0.10870599){\color[rgb]{0,0,0}\makebox(0,0)[lb]{\smash{20}}}%
    \put(0.9838206,0.1842396){\color[rgb]{0,0,0}\makebox(0,0)[lb]{\smash{40}}}%
    \put(0.9838206,0.25977172){\color[rgb]{0,0,0}\makebox(0,0)[lb]{\smash{60}}}%
    \put(0.9838206,0.33530533){\color[rgb]{0,0,0}\makebox(0,0)[lb]{\smash{80}}}%
    \put(0.9838206,0.41083894){\color[rgb]{0,0,0}\makebox(0,0)[lb]{\smash{100}}}%
    \put(0.9838206,0.48637255){\color[rgb]{0,0,0}\makebox(0,0)[lb]{\smash{120}}}%
    \put(0.93544603,0.52981792){\color[rgb]{0,0,0}\makebox(0,0)[lb]{\smash{$\left[{\rm km}^{-1}\right]$}}}%
  \end{picture}%
\endgroup%